\documentclass{amsart}

\usepackage {epsfig}
\usepackage {graphicx}

 \newtheorem{thm}{Theorem}

 \newtheorem{theorem}[thm]{Theorem}
 \theoremstyle{definition}
 
 \theoremstyle{remark}

 \newcommand{\Real}{\mathbb{R}}

\usepackage{graphicx}
\begin{document}

\title[Synchronized oscillations]
{Stability of synchronized oscillations in networks of
phase-oscillators}

\author{Guy Katriel}

\address{}

\email{haggaik@wowmail.com}


\begin{abstract}
We derive simple conditions for the stability or instability of
the synchronized oscillation of a class of networks of coupled
phase-oscillators, which includes many of the systems used in
neural modelling.

\end{abstract}

\maketitle

\section{introduction}

The general equations for a system of $n$ coupled
phase-oscillators are given by
\begin{equation}
\label{gen}
\theta_i'=g_i(\theta_1,\theta_2,...,\theta_n),\;\;\;1\leq i\leq n,
\end{equation}
where $\theta_i$ denotes the phase of the $i$-th oscillator, and
the functions $g_i$ encapsulate both the internal dynamics of the
$i$-th oscillator and its coupling to the other oscillators. Since
the $\theta_i$'s denote phases, we assume that the functions $g_i$
are $2\pi$ periodic with respect to all variables.

The study of the dynamics of networks of phase-oscillators is a
formidable mathematical problem, and it is therefore important to
isolate special structures of such systems which facilitate the
study of at least some aspects of their dynamics. Of course, in
order for the work to be relevant to modelling, these structures
must be natural in the sense that they reflect some modelling
assumptions. In this work we identify a special structure for
systems of the form (\ref{gen}) which makes the study of
synchronization much more tractable. Moreover this structure is
general enough to encompass many of the systems which have been
used in neural modelling in the `frequency domain' approach
\cite{hopp,hibook}.

A {\bf{synchronized oscillation}} of the system (\ref{gen}), is a
solution for which $\theta_i(t)=\overline{\theta}(t)$ for all
$1\leq i \leq n$, with
\begin{equation}
\label{osc} \overline{\theta}(t+T)=\overline{\theta}(t)+2\pi
\;\;\; \forall t\in \Real
\end{equation}
We shall say that the system is {\bf{synchronized}} if there
exists a {\it{stable}} synchronized oscillation (see section
\ref{stabs} for the definition of stability). The study of
synchronization is an important part in the quest to understand
the dynamics of networks of coupled oscillators \cite{pikovsky,
wang}.

We derive simple conditions for the stability/instability of
synchronization for a class of systems. As we shall point out,
these conditions generalize several stability conditions derived
in previous works for more restricted classes of systems.

\vspace{0.5cm}

We first present a special case of the structure that we identify
in this work, before turning to the general case. Assume that the
functions $g_i$ have the special form:
\begin{equation}
\label{spec}
g_i(\theta_1,...,\theta_n)=h(\theta_i)+\sum_{j=1}^{n}{c_{ij}f(\theta_i,\theta_j)},
\end{equation}
Where $h(\alpha)$ is a $2\pi$-periodic function describing the
internal dynamics of the oscillators, $f(\alpha,\beta)$ is a
function which is $2\pi$-periodic in both variables and describes
the coupling of pairs of oscillators, and $c_{ij}$ are constants
describing the strength of the influence of oscillator $j$ on
oscillator $i$. Let us note the modelling assumptions implied by
such a form for the functions $g_i$:

\noindent (i) The influence of all the oscillators on oscillator
$i$ is the sum of terms each one of which represents the influence
of one of the other oscillators. This is known in the literature
as the assumption of {\it{conventional synaptic connections}}.

\noindent (ii) The {\it{type}} of coupling among pairs of
oscillators, represented by the function $f$, is identical for
each pair of oscillators, and only the {\it{strength}} $c_{ij}$ of
the coupling is allowed to vary with $i$ and $j$ (including the
possibility that $c_{ij}=0$, so that there is no influence of
oscillator $j$ on oscillator $i$). A more general model, still
satisfying the assumption of conventional synaptic connections,
would have $f$ depending on the indices $i,j$, but such a model
falls outside the scope of this study and does not admit the
pleasing analytical results proved here.

\noindent (iii) The internal dynamics of the oscillators,
represented by the function $h$, are identical.

Models of the form (\ref{spec}) are already of great interest from
a modelling perspective
\cite{Aria,brown,goel,golumb,hibook,hopp1}, but our results in
fact apply to more general models in which the functions $g_i$
have the form:

\begin{equation}
\label{struct}
g_i(\theta_1,...,\theta_n)=S(\theta_i,\sum_{j=1}^n{c_{ij}f(\theta_i,\theta_j)}).
\end{equation}
The function $S(\phi,y)$, $2\pi$-periodic in the first variable,
can be thought of as a cut-off function limiting the input to each
oscillator (see \cite{hoppe, hopp} for such examples), so its
dependence on $y$ (which we do not restrict) might be sigmoidal or
hump-shaped . We note that this structure relaxes the
`conventional synaptic connections' assumption (i). Systems of the
form (\ref{spec}) are special cases of systems of the form
(\ref{struct}) in which $S(\phi,y)=h(\phi)+y$.

\section{Existence of a synchronized oscillation}

We first deal with the rather trivial matter of existence of a
synchronized oscillation for systems with the structure
(\ref{struct}).

\begin{theorem}
\label{exist} If there exists $c$ such that
\begin{equation}
\label{excond} \sum_{j=1}^n{c_{ij}}=c\;\;\; 1\leq i\leq n
\end{equation}
and if
\begin{equation}
\label{pos} S(\theta,cf(\theta,\theta))>0\;\;\;\forall \theta\in
\Real
\end{equation}
 then there exists a synchronized oscillation
$\theta_1=\theta_2=...=\theta_n=\overline{\theta}$ of
\begin{equation}
\label{msys}
\theta_i'=S(\theta_i,\sum_{j=1}^n{c_{ij}f(\theta_i,\theta_j)}),\;\;\;1\leq
i\leq n
\end{equation}
which is unique up to time-translations.
\end{theorem}

Indeed, the synchronized oscillation is simply the solution of the
scalar equation
\begin{equation}
\label{scalar}
\overline{\theta}'=S(\overline{\theta},cf(\overline{\theta},\overline{\theta})).
\end{equation}

Condition (\ref{excond}) guarantees that each of the equations
(\ref{gen}) reduce to (\ref{scalar}) upon substituting
$\theta_1=\theta_2=...=\theta_n=\overline{\theta}$, while
condition (\ref{pos}) implies that each solution of (\ref{scalar})
satisfies (\ref{osc}) for an appropriate $T>0$. It is easy to see
that (\ref{pos}) is a necessary condition for the existence of a
solution of (\ref{scalar}) satisfying (\ref{osc}), hence for the
existence of a synchronized oscillation. We note that the period
$T$ of the synchronized oscillation is given by the formula
\begin{equation}
\label{period}
T=\int_0^{2\pi}{\frac{d\theta}{S(\theta,cf(\theta,\theta))}}
\end{equation}

We make some remarks about the condition (\ref{excond}):

\noindent (i) Condition (\ref{excond}) says that the sum of
strengths of the couplings of each oscillator to all the others
does not depend on the oscillator, and it is easy to see that if
this condition is violated a synchronized oscillation will usually
not exist, because such a function would have to satisfy two
independent scalar differential equations.

\noindent (ii) Condition (\ref{excond}) can be considered as an
idealization - in a `general' network we do not expect the sum of
strengths of the connections to each oscillator to be exactly
constant. As a consequence, in a `general' network we do not
expect a synchronized oscillation to exist at all. However, there
are two very important classes of systems for which condition will
naturally hold: {\bf{symmetric}} networks and {\bf{regular}}
networks, as we explain below. Moreover, we expect many natural
`large' networks to satisfy (\ref{excond}) in an approximate way,
leading to approximate synchronization - see below.

\noindent (iii) A {\bf{symmetry}} of a network is a permutation
$\tau\in S_n$ with the property that $c_{ij}=c_{\tau(i)\tau(j)}$
for all $1\leq i,j\leq n$. The set of symmetries of the network
forms a subgroup $G$ of $S_n$. It is easy to check that if the
subgroup $G$ is {\it{transitive}}, that is if for any $1\leq
i,j\leq n$, there exists a $\tau\in G$ such that $\tau(i)=j$, then
(\ref{excond}) holds. Thus if we call a network {\bf{symmetric}}
whenever its symmetry group is transitive, we have that any
symmetric network satisfies (\ref{excond}). A simple case is a
cyclically-symmetric ring of oscillators, in which the symmetry
group is generated by the permutation $\tau(i)=i+1\; (mod \;n)$.
For more on symmetry in networks of oscillators see \cite{golu}.

\noindent (iv) We note the trivial fact that a linear combination
of two matrices satisfying (\ref{excond}) also satisfies
(\ref{excond}), so that our condition holds, for example, if we
have two independent sets of connections, each of them with a
different transitive symmetry group.

\noindent (v) Consider the case of a network in which the
connections have either strength $s$ or strength $0$ (i.e., no
connection). Such a network can be represented by a directed graph
in which the oscillators are the nodes and an arrow from node $j$
to node $i$ indicates that $c_{ij}=s$. If this graph is
$k$-{\it{regular}} in the sense that each node has $k$ incoming
arrows, then condition (\ref{excond}) is satisfied. This includes
`random' networks generated by asigning $k$ oscillators to each
oscillator.

\noindent (vi) If we relax the assumption of regularity and
generate a random network by a (perhaps more natural) process of
setting $c_{ij}=s$ with probability $p$ and $c_{ij}=0$ with
probability $1-p$, we can still expect that when the number $n$ of
oscillators is large, the network will satisfy (\ref{excond}) in
an {\it{approximate way}}. Indeed by the central limit theorem the
sums in (\ref{excond}) will be approximately $pns$, with an error
of $O(\frac{1}{\sqrt{n}})$. Thus such a network can be considered
as a small perturbation of a network satisfying (\ref{excond})
with $c=pns$. Under such circumstances, we expect the perturbation
to change the synchronized oscillation into a nearby
{\it{entrained}} oscillation, that is a solution satifying
$\theta_i(t+T)=\theta_i(t)+2\pi$ for all $1\leq i\leq n$. Moreover
we expect the functions $\theta_i$ to be close to one another,
though not identical. We thus have the phenomenon of
`near-synchronization', and we also expect that the stability of
the `near-synchronized' oscillation will be the same as that of
the synchronized solution of the system satisfying (\ref{excond})
off which we perturb. These heuristic remarks need to be backed by
a careful analysis, but we leave this to future work.
\vspace{0.5cm}

We formulate the corollary of theorem \ref{exist} applied to
systems of the form (\ref{spec}):

\begin{theorem}
\label{exist1}  Assume (\ref{excond}) holds and
\begin{equation}
\label{pos1} h(\theta)+cf(\theta,\theta)>0\;\;\;\forall \theta\in
\Real
\end{equation}
then there exists a synchronized oscillation
$\theta_1=\theta_2=...=\theta_n=\overline{\theta}$ of
\begin{equation}
\label{conventional}
\theta_i'=h(\theta_i)+\sum_{j=1}^{n}{c_{ij}f(\theta_i,\theta_j)},\;\;\;1\leq
i\leq n
\end{equation}
which is unique up to time-translations.
\end{theorem}

\section{stability of the synchronized oscillation}
\label{stabs}

We now turn to the more interesting issue of stability of the
synchronized oscillation.

\label{stable} We recall some standard definitions. Let $d$ denote
the euclidean distance in $\Real^n$. A solution
$\theta(t)=(\theta_1(t),...,\theta_n(t))$ of (\ref{gen}) is said
to be {\bf{Liapunov stable}} if for any $\epsilon>0$ there is a
$\delta>0$ such that whenever $t_0\in
\Real,\;{\tilde{\theta}}_0\in \Real^n$ and
$d({\tilde{\theta}}_0,\theta(t_0))<\delta$, we have
$d({\tilde{\theta}}(t),\theta(t))<\epsilon$ for all $t>0$, where
${\tilde{\theta}}(t)$ denotes the solution of (\ref{gen}) with
initial condition ${\tilde{\theta}}(0)={\tilde{\theta}}_0$.

$\theta(t)$ is said to be (asymptotically) {\bf{stable}} if it is
Liapunov stable {\it{and}} there exists a $\delta_1>0$ such that
whenever $t_0\in \Real,\;{\tilde{\theta}}_0\in \Real^n$ and
$d({\tilde{\theta}}_0,\theta(t_0))<\delta_1$, there exists some
$t_1\in\Real$ such that
$\lim_{t\rightarrow\infty}{d({\tilde{\theta}}(t),\theta(t-t_1))}=0$,
where ${\tilde{\theta}}(t)$ denotes the solution of (\ref{gen})
with initial condition ${\tilde{\theta}}(0)={\tilde{\theta}}_0$.

$\theta(t)$ is said to be {\bf{unstable}} if it is {\it{not}}
Liapunov stable.

We note that we are dealing here with {\it{local}} stability, so
that the fact that the synchronized oscillation is stable does not
imply that synchronization will be achieved from any initial
state, but only that there is some open set of initial conditions
that will lead to synchronization. The study of global
synchronization properties of systems of phase-oscillators is a
very intriguing problem.

In general the stability of a synchronized oscillation is to be
determined by examining the time-periodic linear system obtained
by linearizing (\ref{gen}) around the synchronized solution:
\begin{equation}
\label{lin} u'=A(t)u ,\;\;\;1\leq i\leq n,
\end{equation}
where $A(t)$ is the $n\times n$ matrix with
\begin{equation}
\label{line}A_{ik}(t)= \frac{\partial g_i}{\partial
\theta_k}(\overline{\theta}(t),...,\overline{\theta}(t)),
\end{equation}
 which in general can
only be done by numerical computation. The crucial point is that
the special structure (\ref{struct}) entails a special structure
for the linearized system (\ref{lin}), which makes it possible to
proceed much further with the analytic investigation than in the
general case. Differentiating (\ref{struct}) we have
\begin{eqnarray}
\label{pardif} \frac{\partial g_i}{\partial
\theta_k}(\theta_1,...,\theta_n)&=&\delta_{ik}\frac{\partial
S}{\partial
\phi}(\theta_i,\sum_{j=1}^n{c_{ij}f(\theta_i,\theta_j)})\\&+&
c_{ik}\frac{\partial S}{\partial
y}(\theta_i,\sum_{j=1}^n{c_{ij}f(\theta_i,\theta_j)})\frac{\partial
f}{\partial \beta}(\theta_i,\theta_k)
\nonumber\\
&+&\delta_{ik} \frac{\partial S}{\partial
y}(\theta_i,\sum_{j=1}^n{c_{ij}f(\theta_i,\theta_j)})\sum_{j=1}^n{c_{ij}\frac{\partial
f}{\partial \alpha}(\theta_i,\theta_j)}\nonumber.
\end{eqnarray}
Thus substituting
$\theta_1=\theta_2=...=\theta_n=\overline{\theta}$ (\ref{pardif})
we obtain
\begin{equation}
\label{forma}
A(t)=a(\overline{\theta}(t))I+b(\overline{\theta}(t))C,
\end{equation}
 where $I$ is the
identity matrix, $C$ is the {\it{connection matrix}}
$C_{ij}=c_{ij}$, and the scalar functions $a(\theta),b(\theta)$
are given by:
\begin{equation} \label{a}a(\theta)=\frac{\partial
S}{\partial \phi} (\theta,cf (\theta ,\theta))+c\frac{\partial
S}{\partial y}(\theta,cf(\theta, \theta))\frac{\partial
f}{\partial \alpha} (\theta,\theta)
\end{equation}
\begin{equation}
\label{b} b(\theta)=\frac{\partial S}{\partial
y}(\theta,cf(\theta,\theta))\frac{\partial f}{\partial
\beta}(\theta,\theta).
\end{equation}

Let us assume for the moment that the connection matrix $C$ is
diagonalizable:
\begin{equation}
\label{diag} C=M^{-1}DM
\end{equation}
with $D$ the diagonal matrix
$$D=diag[\lambda_1,...,\lambda_n],$$
where $\lambda_k$, $1\leq k \leq n$ are the eigenvalues of $C$. We
note that (\ref{excond}) implies that $(1,1,...,1)$ is an
eigenvector of $C$ with eigenvalue $c$, so that we may assume
\begin{equation}
\label{lambda1} \lambda_1=c.
\end{equation}

Let $F(t)$  denote the $n\times n$ matrix-valued function which is
the fundamental solution of (\ref{lin}), so that $F(0)=I$ and
\begin{equation}
\label{fund} F'=A(t)F. \end{equation}
We recall from the stability
theory for periodic solutions of autonomous systems  that the
synchronized oscillation is stable if all the eigenvalues of
$F(T)$ ($T$ denoting the period of $\overline{\theta}$), known as
the {\it{characteristic multipliers}}, are in the interior of the
unit disk in the complex plane, except for one eigenvalue which is
$1$ (related to time-translation invariance). If at least one of
the eigenvalues is in the exterior of the unit disk, the
synchronized oscillation is unstable (see, {\it{e.g.}},
\cite{robinson}, Ch. V, theorem 8.4).

Defining $G=MFM^{-1}$, we have from (\ref{forma}), (\ref{fund})
\begin{equation}
\label{ds}
G'=[a(\overline{\theta}(t))I+b(\overline{\theta}(t))D]G,
\end{equation}
with $G(0)=I$.  Since the linear system (\ref{ds}) is diagonal,
its solution is
$$G(t)=diag[r_1(t),...,r_n(t)]$$
where the functions $r_i$ satisfy the scalar equations
\begin{equation}
\label{ri} r_i'(t)=[a(\overline{\theta}(t))+\lambda_i
b(\overline{\theta}(t))]r_i(t),
\end{equation}
 with $r_i(0)=1$, whose solution is
$$r_i(t)=e^{\int_0^t{[a(\overline{\theta}(s))+\lambda_i b(\overline{\theta}(s))]ds}}.$$
In particular we have
$$G(T)=diag[e^{\int_0^T{[a(\overline{\theta}(t))+\lambda_1 b(\overline{\theta}(t))]dt}},...,e^{\int_0^T{[a(\overline{\theta}(t))+\lambda_n
b(\overline{\theta}(t))]dt}}],$$ so that the eigenvalues of $F(T)$
are
\begin{equation}
\label{mu} \mu_i=e^{\sigma_i},\;\;\;1\leq i \leq n.
\end{equation}
where
\begin{equation}
\label{defsing}
\sigma_i=\int_0^T{[a(\overline{\theta}(t))+\lambda_i
b(\overline{\theta}(t))]dt}
\end{equation}

By making the change of variable $\theta=\overline{\theta}(t)$, so
that, using (\ref{scalar}),
$d\theta=S(\theta,cf(\theta,\theta))dt$, we obtain
\begin{equation}
\label{chint} \sigma_i=\int_0^{2\pi}{[a(\theta)+\lambda_i
b(\theta)]\frac{d\theta}{S(\theta,cf(\theta,\theta))}}.
\end{equation}
In particular, for $i=1$, using (\ref{lambda1}),
\begin{equation}
\label{sig}\sigma_1=\int_0^{2\pi}{[a(\theta)+c
b(\theta)]\frac{d\theta}{S(\theta,cf(\theta,\theta))}}
\end{equation}
and using the easily-verified fact that
$$\frac{d}{d\theta}[S(\theta,cf(\theta,\theta)]=a(\theta)+cb(\theta),$$
we obtain
$$\sigma_1=\int_0^{2\pi}{\frac{d}{d\theta}[S(\theta,cf(\theta,\theta))]
}\frac{d\theta}{S(\theta,cf(\theta,\theta))}=
\int_0^{2\pi}{\frac{d}{d\theta}[\log(S(\theta,cf(\theta,\theta)))]d\theta}=0.
$$
Thus $\mu_1=1$ is the eigenvalue of $F(T)$ corresponding to
time-translations, and the sufficient condition for stability of
the synchronized solution is that all the other eigenvalues, given
by (\ref{mu}), are inside the unit disc. This is equivalent to the
condition that the numbers $\sigma_i$ have {\it{negative real
parts}} for $2\leq i\leq n$.

We can express $\sigma_i$ in a somewhat simpler form by noting
that from (\ref{sig}) and the fact that $\sigma_1=0$ we have
$$\int_0^{2\pi}{
\frac{a(\theta) d\theta}{S(\theta,cf(\theta,\theta))}}
=-c\int_0^{2\pi}{\frac{b(\theta)
d\theta}{S(\theta,cf(\theta,\theta))}}$$
so we can rewrite
(\ref{chint}) as
$$\sigma_i=(\lambda_i-c)\int_0^{2\pi}{
\frac{b(\theta) d\theta}{S(\theta,cf(\theta,\theta))}}.$$
Recalling the definition (\ref{b}) of $b(\theta)$, we finally
obtain the expression:
\begin{equation}
\label{for}
 \sigma_i=(\lambda_i-c)\int_0^{2\pi}{\frac{\partial
S}{\partial y}(\theta,cf(\theta,\theta))\frac{\partial f}{\partial
\beta}(\theta,\theta)
\frac{d\theta}{S(\theta,cf(\theta,\theta))}},\;\;\;1\leq i \leq n.
\end{equation}

We note now that although (\ref{for}) was derived under the
assumption that $C$ is diagonalizable, it continues to be valid
for any matrix $C$ satisfying (\ref{excond}), because the set of
diagonalizable matrices is dense and both sides of (\ref{for}) are
continuous in $C$. We thus have:

\begin{theorem}
\label{main1} Assume (\ref{excond}) and (\ref{pos}) are satisfied.
Let \begin{equation} \label{defchi}
\chi=\chi(S,f,c)=\int_0^{2\pi}{\frac{\partial S}{\partial
y}(\theta,cf(\theta,\theta))\frac{\partial f}{\partial
\beta}(\theta,\theta)
\frac{d\theta}{S(\theta,cf(\theta,\theta))}}.
\end{equation}

\noindent (i) If
\begin{equation}
\label{stc} \chi[Re(\lambda_i)-c]<0,\;\;\;2\leq i\leq n
\end{equation}
then the synchronized oscillation of (\ref{msys}) is stable.

\noindent (ii) If, for some $2\leq i\leq n$,
$\chi[Re(\lambda_i)-c]>0$ then the synchronized oscillation of
(\ref{msys}) is unstable.
\end{theorem}

We formulate the stability result for the special case of
conventional synaptic coupling:

\begin{theorem}
\label{main2} Assume (\ref{excond}) and (\ref{pos1}) are
satisfied. Let
\begin{equation}
\label{defchi1} \chi=\chi(h,f,c)=\int_0^{2\pi}{\frac{\partial
f}{\partial \beta}(\theta,\theta)
\frac{d\theta}{h(\theta)+cf(\theta,\theta)}}.
\end{equation}

\noindent (i) If (\ref{stc}) holds then the synchronized
oscillation of (\ref{conventional}) is stable.

\noindent (ii) If, for some $2\leq i\leq n$,
$\chi[Re(\lambda_i)-c]>0$ then the synchronized oscillation of
(\ref{conventional}) is unstable.
\end{theorem}

The calculations involved in determining the stability of the
synchronized oscillation using theorems \ref{main1} or \ref{main2}
split into two parts: calculation of the integral $\chi$, and
study of the spectrum of the connection matrix $C$. The following
reformulation of theorem \ref{main1} highlights the independence
of these two aspects:

\begin{theorem}
\label{main3} Assume (\ref{excond}),(\ref{pos}) hold, and let
$\chi$ be defined by (\ref{defchi}).

(i) If
\begin{equation}
\label{neg}\min_{2\leq i\leq n}{Re(\lambda_i)}<c<\max_{2\leq i\leq
n}{Re(\lambda_i)} \end{equation} then the synchronized solution of
any system of the form (\ref{msys}) with $\chi\neq 0$ is unstable.

\noindent (ii) If
\begin{equation}
\label{p1}\min_{2\leq i\leq n}{Re(\lambda_i)}>c
\end{equation}
then the synchronized solution of (\ref{msys}) is stable if
$\chi<0$ and unstable if $\chi>0$.

\noindent (iii) If
\begin{equation}
\label{p2}\max_{2\leq i\leq n}{Re(\lambda_i)}<c
 \end{equation}
then the synchronized solution of (\ref{msys}) is stable if
$\chi>0$ and unstable if $\chi<0$.
\end{theorem}

We note that $\chi$ depends on the connection matrix only through
the `overall strength' $c$.

Let us study, as an example, the system
\begin{equation}
\label{example}
\theta_i'=\omega-\sin(\theta_i)\sum_{j=1}^n{c_{ij}\cos(\theta_j)},
\end{equation}
considered in \cite{hopp1}. This is a system of the type
(\ref{spec}), with
$$h(\alpha)=\omega,\;\;\;f(\alpha,\beta)=-\sin(\alpha)\cos(\beta).$$
We assume (\ref{excond}) holds. In order to satisfy (\ref{pos1})
we must have $\omega-c\sin(\theta)\cos(\theta)>0$ for all
$\theta$, which is easily seen to be equivalent to the requirement
\begin{equation}
\label{pexample} \omega> \frac{|c|}{2}.
\end{equation}
Thus if (\ref{excond}) and {\ref{pexample}) hold, there exists a
synchronized oscillation of (\ref{example}) computing $\chi$
according to (\ref{defchi1}) we obtain
$$\chi=\int_0^{2\pi}{\sin^2(\theta)
\frac{d\theta}{\omega-c\sin(\theta)\cos(\theta)}},$$ and we see
that $\chi>0$ whenever (\ref{pexample}) holds. Thus from theorem
\ref{main2} we obtain that the synchronized solution is stable iff
the real parts of the eigenvalues of $C$ (other than $c$ itself)
are all smaller than $c$, or, in other words, when all the
eigenvalues of $C-cI$ have negative real parts.

\section{Connection matrices and the stability of synchronized oscillations}

The spectrum of the connection matrix $C$ is, of course, totally
independent of the nonlinearities $h,f,S$. This fact allows us,
for some connection matrices, to conclude that the synchronized
solution is unstable, independently of the internal dynamics of
the oscillators and of the coupling functions, as shown by part
(i) of theorem \ref{main3}.

In this section we consider the role of the connection matrix
through some examples, using some well-known results from linear
algebra.

\subsection{Two-oscillator networks}

In the case $n=2$ it is easy to see that if we assume no
self-coupling, ($c_{11}=c_{22}=0$) the general system (\ref{msys})
is in fact no more general than the particular case
(\ref{conventional}), and in this case we may also absorb $h$ into
the $f$, so we are dealing with the system:
\begin{eqnarray}
\label{two}
\theta_1'=f(\theta_1,\theta_2)\\
\theta_2'=f(\theta_2,\theta_1) \nonumber
\end{eqnarray}
The connection matrix is
$$C=\begin{pmatrix}
  0 & 1 \\
  1 & 0
\end{pmatrix}$$
and we have $\lambda_1=c=1, \lambda_2=-1$  Thus (\ref{p2}) holds
and from theorem \ref{main3} we obtain:

\begin{theorem}
For a system of two coupled phase-oscillators of the form
(\ref{two}) for which
$$f(\theta,\theta)>0\;\;\;\forall \theta\in\Real,$$
let:
$$\chi=\int_0^{2\pi}\frac{\partial f}{\partial
\beta}(\theta,\theta)\frac{d\theta}{f(\theta,\theta)}.$$ The
synchronized solution is stable if $\chi>0$ and unstable if
$\chi<0$.
\end{theorem}

\subsection{Three-oscillator networks}

We assume that there is no self-coupling, so that the connection
matrix can be written in the form
\begin{equation}
\label{matrix3}
 C=\begin{pmatrix}
  0 & c_{12} & c-c_{12} \\
  c-c_{23} & 0 & c_{23} \\
  c_{31} & c-c_{31} & 0
\end{pmatrix}.
\end{equation}
By computing the characteristic polynomial of $C-cI$, we find that
the product of its two eigenvalues (besides the eigenvalue $0$) is
$$(\lambda_2-c)(\lambda_3-c)=\Delta$$
where
\begin{equation}
\label{delta}
\Delta=3c^2-(c_{12}+c_{23}+c_{31})c+(c_{12}c_{23}+c_{12}c_{31}+c_{23}c_{31}),
\end{equation}
and that their sum is
$$(\lambda_2-c)+(\lambda_3-c)=-3c.$$

Assume $\Delta<0$. Then $\lambda_2$,$\lambda_3$ must be real
(because otherwise they are conjugate which would imply that the
product $(\lambda_2-c)(\lambda_3-c)$ is non-negative), and
$(\lambda_2-c)$, $(\lambda_3-c)$ are of opposite signs, hence part
(i) of theorem \ref{main3} implies that the synchronized
oscillation is unstable.

Now assume $\Delta>0$. If $\lambda_2$,$\lambda_3$ are non-real,
hence conjugate, then we have
$$Re(\lambda_2)-c=Re(\lambda_3)-c=-\frac{3}{2}c$$
so that (\ref{p1}) holds if $c<0$ and (\ref{p2}) holds if $c>0$.
If $\lambda_2, \lambda_3$ are real then $\Delta>0$ implies that
$(\lambda_2-c)$, $(\lambda_3-c)$ are of the same sign, which is
opposite to the sign of $c$, so that again (\ref{p1}) holds if
$c<0$ and (\ref{p2}) holds if $c>0$.

We thus obtain:
\begin{theorem}
\label{three} Consider the system (\ref{msys}) with $n=3$ and
connection matrix given by (\ref{matrix3}), and assume that
(\ref{pos}) holds. Define $\chi$ by (\ref{defchi}).

\noindent (i) If $\Delta<0$ and $\chi\neq 0$ (where $\Delta$ is
given by (\ref{delta})) then the synchronized oscillation is
unstable.

\noindent (ii) If $\Delta>0$ then the synchronized oscillation is
stable if $c\chi>0$ and unstable if $c\chi<0$.
\end{theorem}

\subsection{Larger networks} In principle, for larger networks,
 with theorem
\ref{main3}, one can obtain algebraic conditions involving the
elements of the connection matrix and $\chi$ for determining the
stability/instability of the synchronized solution, by applying
the Routh-Hurwitz conditions to the characteristic polynomial of
$C-cI$. Obviously these conditions become more complicated as the
size of the network gets larger, and we shall not proceed along
these lines but rather consider some specific classes of
connection matrices for which simple results are possible.

\subsection{Non-negative connection matrices}
A very important case is that in which all the entries of the
connection matrix are non-negative. In this case we can apply the
Perron-Frobenius theorem to obtain stability results. In fact to
apply the Perron-Frobenius theorem we need to assume also that the
matrix $C$ is irreducible, which means that there does not exist a
partition of the indices $\{1,...,n\}$ into two disjoint sets
$A,B$ such that $i\in A$, $j\in B$ implies $c_{ij}=0$. In other
words, we assume that we cannot partition the oscillators into
disjoint sets $A,B$ so that oscillators in $B$ have no influence
on oscillators in $A$ (we note that a simple sufficient condition
for $C$ to be irreducible is that $c_{ij}\neq 0$ for all $i\neq
j$). The Perron-Frobenius theorem states that if $C$ is
non-negative and irreducible then it has a unique eigenvalue
$\lambda$ which is positive and has an eigenvector whose
components are positive. Moreover $\lambda$ is a simple eigenvalue
and all other eigenvalues of $C$ are less than or equal in modulus
to $\lambda$.

Since in our case we already know of the eigenvalue $\lambda_1=c$
which has an eigenvector $(1,1,...,1)$ with positive components,
the uniqueness part of the Perron-Frobenius theorem implies that
all other eigenvalues satisfy $\lambda_i\neq c$ and
$|\lambda_i|\leq c$. These two facts imply that $Re(\lambda_i)<c$,
so that (\ref{p2}) holds. We thus conclude:

\begin{theorem}
\label{nonneg} If the connection matrix $C$ satisfies
(\ref{excond}) and is non-negative and irreducible, and
(\ref{pos}) holds, then the synchronized solution of (\ref{msys})
is stable if $\chi>0$ and unstable if $\chi<0$, where $\chi$ is
defined by (\ref{defchi}).
\end{theorem}

Of course an analogous theorem is true for the case of
non-positive matrices, where the roles of the conditions $\chi>0$,
$\chi<0$ are reversed.

A special case of theorem \ref{nonneg} was proved in \cite{goel}
(section 2.5), with $\chi$ given in a somewhat less explicit form.
There it was assumed that the system is of the form
(\ref{conventional}) with $h$ a constant and that $f$ is of the
special product form $f(\alpha,\beta)=f_1(\alpha)f_2(\beta)$. The
unneccesary assumption that $C$ is a symmetric matrix was made.

We note that theorem \ref{nonneg} includes the case of global
uniform coupling given by $c_{ij}=s$ for all $i,j$ (or for all
$i\neq j$). This case has been treated, for the special case of
(\ref{conventional}) in which the coupling $f(\alpha,\beta)$
depends only $\beta$, in \cite{golumb}, and for the special case
in which the coupling is of the form
$f(\alpha,\beta)=f_0(\beta-\alpha)+f_1(\alpha)f_2(\beta)$ and $h$
is constant, in \cite{brown}. The stability results for
synchronized oscillations in these works are corollaries of
theorem \ref{nonneg}.

\subsection{Cyclically symmetric systems}
We now consider the case of a ring of oscillators with cyclic
symmetry, which means that the effect of the $i$-th oscillator on
the $j$-th oscillator is the same as that of the $i+k$-th
oscillator $j+k$-th oscillator (where $i+k$, $j+k$ are considered
modulo $n$). In mathematical terms this means that the connection
matrix $C$ is {\it{circulant}} \cite{davis}, that is, one having
the form
\begin{equation}
\label{cyc} c_{ij}=d_{(j-i+1)\; mod\; n},
\end{equation}
with $d\in \Real^n$  some vector. As we noted before, a circulant
matrix automatically satisfies (\ref{excond}).

We have the following formula for the eigenvalues of a circulant
matrix, where $\rho =e^{-2\pi i/n}$:
\begin{equation}
\label{eigs} \lambda_k=\sum_{m=1}^{n}{d_m\rho^{(k-1)(m-1)}}.
\end{equation}
This implies
\begin{equation}
\label{cycfor} Re(\lambda_k)=\sum_{m=1}^{n}{d_m \cos \Big(2\pi
\frac{(k-1)(m-1)}{n}\Big)}.
\end{equation}

The formula (\ref{cycfor}), together with the identity
$1-\cos(x)=2\sin^2(\frac{x}{2})$, give us the following stability
result for cyclically symmetric matrices:

\begin{theorem}
\label{cc} Assume (\ref{pos}) is satisfied, and assume that the
network is cyclically symmetric, its connection matrix given by
(\ref{cyc}). Let $\chi$ be defined by (\ref{defchi}), and define,
for $1\leq k \leq n$:
$$\gamma_k=\sum_{m=1}^{n}{d_m \sin^2 \Big(\pi
\frac{(k-1)(m-1)}{n}\Big)}.$$

\noindent (i) If some of the $\gamma_k$'s are postitive and some
are negative, the synchronized oscillation of (\ref{msys}) is
unstable.

\noindent (ii) If $\chi \gamma_k>0$ for all $1\leq k \leq n$ then
the synchronized oscillation is stable.

\noindent (iii) If $\chi \gamma_k<0$ for all $1\leq k \leq n$ then
the synchronized oscillation is unstable.

\end{theorem}

\subsection{Two populations of interacting oscillators}

Let us consider $2n$ oscillators, subdivided into two groups, each
of size $n$, with the pairs of oscillators within each group
coupled with strength $p$, and any pair of oscillators belonging
to different groups are coupled with strength $q$.

\begin{eqnarray*}
c_{ij}&=&p\;\;if\; 1\leq i\neq j \leq n\;or\; n+1\leq i\neq j\leq 2n\\
&=&q\;\;if\; 1\leq i \leq n,\; n+1\leq j\leq 2n,\\
&=&q\;\;if\; n+1\leq i \leq 2n,\; 1\leq j\leq n,\\
&=&0 \;\;if \;\;i=j.
\end{eqnarray*}

For example in the case $n=3$ the connection matrix is
$$C=
\begin{pmatrix}
  0 & p & p & q & q & q \\
  p & 0 & p & q & q & q \\
  p & p & 0 & q & q & q \\
  q & q & q & 0 & p & p \\
  q & q & q & p & 0 & p \\
  q & q & q & p & p & 0 \\
\end{pmatrix}
$$

Such a matrix satisfies (\ref{excond}) with $c=(n-1)p+nq$. Let us
denote by $e_k\in \Real^{2n}$ ($1\leq k \leq 2n$) vector with $1$
in the $k$-th place and $0$ elesewhere. It is is easy to verify
that the eigenvalues of $C$ are $c$ with eigenvector
$(1,1,1...,1)$, $(n-1)p-nq$ with eigenvector
$(1,1,..1,-1,-1,...-1)$, and $-p$ with the eigenvectors $v_i$,
$w_i$ ($2\leq i\leq n$), given by
$$
v_{i}=e_1 - e_i,\;\;\;2\leq i\leq n, \\
$$
$$
w_{i}=e_{n+1} - e_{n+i},\;\;\;2\leq i\leq n.
$$
We thus see that (\ref{p1}) holds if $p+q>0$ and $q>0$, (\ref{p2})
holds if $p+q<0$ and $q<0$ and (\ref{neg}) holds if $q(p+q)<0$. We
thus obtain:

\begin{theorem}
Consider the system:
\begin{eqnarray*}
\theta_i'&=&S\Big(\theta_i,p\sum_{{\tiny{\begin{array}{c}j=1\\
j\neq i\end{array}}}}^n{f(\theta_i,\theta_j)}+q\sum_{j=n+1}^{2n}{f(\theta_i,\theta_j)}\Big),\;\;\;1\leq i\leq n\\
\theta_i'&=&S\Big(\theta_i,q\sum_{j=1}^n{f(\theta_i,\theta_j)}+p\sum_{{\tiny{\begin{array}{c}j=n+1\\
j\neq i
\end{array}}}}^{2n}{f(\theta_i,\theta_j)}\Big),\;\;\;n+1\leq i\leq
2n
\end{eqnarray*}
and assume (\ref{excond}) holds. Define $\chi$ by (\ref{defchi}).

 \noindent (i) If
$q(p+q)<0$ and $\chi\neq 0$ the synchronized oscillation is
unstable.

\noindent (ii) If $p+q<0$ and $q<0$ then the synchronized
oscillation is stable if $\chi<0$ and stable if $\chi>0$.

\noindent (iii) If $p+q>0$ and $q>0$ then the synchronized
oscillation is stable if $\chi>0$ and unstable if $\chi<0$.
\end{theorem}


\begin{thebibliography}{9}

\bibitem{Aria} J.T. Ariaratnam \& S.H. Strogatz,
{\it{Phase Diagram for the Winfree model of coupled nonlinear
oscillators}}, Phys. Rev. Letters {\bf{86}} (2001), 4278-4281.

\bibitem{brown} E. Brown, P. Holmes \& J. Moehlis, {\it{Globally coupled oscillator networks}},
In: Perspectives and Problems in Nonlinear Science: A Celebratory
Volume in Honor of Larry Sirovich, Springer-Verlag (New York),
2003

\bibitem{davis} P.J. Davis, `Circulant Matrices', Chelsea
(New-York), 1997.

\bibitem{goel} P. Goel \& B. Ermentrout,
{\it{Synchrony, stability, and firing patterns in pulse-coupled
oscillators}}, Physica D {\bf{163}} (2002), 191-216.

\bibitem{golumb} D. Golumb, D. Hansel, B. Shraiman \& H.
Sompolinsky, {\it{Clustering in globally coupled phase
oscillators}}, Phys. Rev. A {\bf{45}} (1992), 3516-3530.

\bibitem{golu} M. Golubitsky \& I. Stewart, {\it{Patterns of
oscillation in coupled cell systems}},
 in Geometry, Dynamics and Mechanics: 60th Birthday Volume
 for J.E. Marsden (eds. P.Holmes, P.Newton, and A.Weinstein),
  Springer-Verlag (New-York), 2002.

 \bibitem{hoppe} F.C. Hoppensteadt, {\it{Signal processing by
 model neural networks}}, SIAM Rev. {\bf{34}} (1992), 426-444.


\bibitem{hopp} F.C. Hoppensteadt,
 `An Introduction to the Mathematics of Neurons', Cambridge University Press (Cambridge), 1997.


\bibitem{hibook} F.C. Hoppensteadt \& E.M. Izhikevich,
 `Weakly Connected Neural Networks', Springer-Verlag (New-York), 1997.

\bibitem{hopp1}  F.C. Hoppensteadt \& E.M. Izhikevich,
{\it{Pattern recognition via synchronization in phase-locked loop
neural networks}}, IEEE Trans. Neural Networks {\bf{11}} (2000),
734-738.

\bibitem{pikovsky} A. Pikovsky, M. Rosenblum \& J. Kurths, `Synchronization',
Cambridge University Press (Cambridge), 2001.

\bibitem{robinson} C. Robinson, `Dynamical Systems', CRC Press
(Boca Raton), 1995.


\bibitem{wang}  X.F. Wang,
{\it{Complex networks: topology, dynamics and synchronization}},
Int. J. Bifurcations and Chaos {\bf{12}} (2002), 885-916.

\end{thebibliography}
\end{document}